\documentclass[12pt]{article}
\usepackage{amsmath, amsthm, amssymb, amstext}
\usepackage{mathrsfs}
\usepackage{array, amsfonts, mathrsfs}
\usepackage{latexsym, amsmath}
\usepackage{graphicx,color}
\usepackage{epsf}

\newtheorem{Lemma}{Lemma}
\newtheorem{Theorem}{Theorem}

\newtheorem{Corollary}{Corollary}

\date{}

\begin{document}

\author{M.I.Belishev\thanks {St. Petersburg Department of Steklov Mathematical
        Institute, St.Petersburg, Russia, e-mail: belishev@pdmi.ras.ru. Supported
        by the RFBR grant 20-01 627-À and Volks-Wagen Foundation.},\,
        D.V.Korikov\thanks {St.Petersburg Department of Steklov Mathematical
        Institute, St. Petersburg, Russia, e-mail: thecakeisalie@list.ru. Supported
        by  Russian Science Foundation, grant No. 17-11-01126}.}

\title{On the EIT problem for nonorientable surfaces}
\maketitle

\begin{abstract}
Let $(\Omega,g)$ be a smooth compact two-dimensional Riemannian
manifold with boundary, $\Lambda_g: f\mapsto
\partial_\nu u|_{\partial\Omega}$ its DN map, where $u$ obeys $\Delta_g
u=0$ in $\Omega$ and $u|_{\partial \Omega}=f$. The Electric
Impedance Tomography problem is to determine $\Omega$ from
$\Lambda_g$.

A criterion is proposed that enables one to detect (via
$\Lambda_g$) whether $\Omega$ is orientable or not.

The algebraic version of the BC-method is applied to solve the EIT
problem for the Moebius band. The main instrument is the algebra
of holomorphic functions on the double covering ${\mathbb M}$ of
$M$, which is determined by $\Lambda_g$ up to an isometric
isomorphism. Its Gelfand spectrum (the set of characters) plays
the role of the material for constructing a relevant copy
$(M',g')$ of $(M,g)$. This copy is conformally equivalent to the
original, provides $\partial M'=\partial
M,\,\,\Lambda_{g'}=\Lambda_g$, and thus solves the problem.
\end{abstract}


\bigskip

\section{Introduction}\label{sec Introduction}

\subsubsection*{About the paper}
The fact that the Dirichlet-to-Neumann map does determine the
Riemannian surface with boundary up to conformal equivalence, is
well known \cite{LUEns,BCald,HMich}. It was first established in
\cite{LUEns}. In \cite{HMich} the explicit complex analysis
formulas for determination of the (oriented) Riemannian surface
are provided.

In \cite{BCald} this fact is proved for the (oriented) Riemannian
surfaces by the use of the connection between the EIT problem and
Banach algebras of analytic functions. Our prospective goal is to
extend the approach \cite{BCald} to the {\it nonorientable}
surfaces. The present paper is the first step in this direction.

\subsubsection*{Results}
\noindent$\bullet$\,\,\,Let $(\Omega,g)$ be a two-dimensional
smooth\footnote{everywhere in the paper, {\it smooth} means
$C^\infty$-smooth} compact Riemannian manifold endowed with the
smooth metric tensor $g$, $\Delta_g$ the Beltrami-Laplace operator
on $M$. Let $u=u^f(x)$ be a solution to the elliptic Dirichlet
boundary value problem
 \begin{align}
\label{Eq Laplace} & \Delta_g u = 0 && \text{in}\,\,{\rm int\,}\Omega\\
\label{Eq Dir data} & u=f && \text{on}\,\,\Gamma,
 \end{align}
where ${\rm int\,}\Omega:=\Omega\setminus\partial\Omega$ and
$\partial \Omega=:\Gamma=\Gamma_1\cup\dots\cup\Gamma_N$. The
Dirichlet-to-Neumann map of $\Omega$ is the operator $\Lambda_g:
f\mapsto
\partial_\nu u^f|_\Gamma$, where $\nu$ is the outward normal to
$\Gamma$.

Our main result is Theorem \ref{Th Criterion} that provides a
criterion, which enables one to detect orientability of the
surface $\Omega$ via its DN map $\Lambda_g$. We prove that
orientability is equivalent to the solvability of a Hamilton-type
system on $\Gamma$ with the `Hamiltonian' $\Lambda_g$. If the
boundary consists of the single component, the criterion is
simplified: $\Omega$ is orientable iff ${\rm Ker\,}[I\!+\!(\Lambda
J)^2]\not=\{0\}$, where $J$ is the integration along $\Gamma$. It
is noteworthy that the operator $I\!+\!(\Lambda J)^2$ introduced
in \cite{BCald} has multidimensional analogs \cite{B CUBO 3d
tomogr,BSharaf}.
\smallskip

\noindent$\bullet$\,\,\,The possibilities of the algebraic
approach \cite{BCald} for nonorientable surfaces are demonstrated
by the example of EIT problem for the Moebius band $(M,g)$. The
main tool for solving is the algebra of the boundary values of the
holomorphic functions on the (orientable) double covering
${\mathbb M}$ of $M$. This algebra is determined by $\Lambda_g$ up
to isometry. Its Gelfand spectrum (the set of characters) plays
the role of the material for constructing a relevant copy
$(M',g')$ of $(M,g)$. This copy is conformally equivalent to the
original, provides $\partial M'=\partial
M,\,\,\Lambda_{g'}=\Lambda_g$, and thus solves the problem. Note
that to construct such a copy is the only relevant understanding
of `to solve the EIT problem for the {\it unknown} manifold'
\cite{BCald,B Sobolev Geom Rings,B UMN}.

\subsubsection*{Comments}

\noindent$\bullet$\,\,\,According to the physical meaning of EIT
problem, an external observer must reconstruct the shape of the
conducting shell $\Omega$ from measurements taken at its border
$\Gamma$. The observer prospects the shell with electric current
$\nabla_g u^f$ initiated by potential $f$ applied to the border,
and registers the current $\partial_\nu u^f=\Lambda_g f$ flowing
across the border. The above mentioned criterion enables him to
determine whether the shell is orientable or not, {\it without
solving the inverse problem}.
\smallskip

\noindent$\bullet$\,\,\,In the most general case, the shell has a
multicomponent boundary and is homeomorphic to a sphere with
handles and Mobius bands glued into it.  To visualize such a
complex structure from the boundary is a worthy and challenging
task.

\subsubsection*{Acknowledgements}
The authors are very much grateful to A.F.Vakulenko for useful
discussions on the subject of the paper. His advices allowed us to
significantly simplify the proof of the basic Theorem 1.

\section{Orientability via DN-map}\label{sec Orientability via DN-map}

\subsubsection*{Harmonic functions}

\noindent$\bullet$\,\,\,In the sequel, $(\Omega,g)$ is a 2d smooth
compact Rimannian manifold with the boundary
$\Gamma=\Gamma_1\cup\dots\cup\Gamma_N$, where $\Gamma_j$ are
diffeomorphic to the circle in $\mathbb R^2$. For
$A\subset\Omega$, we denote
 $$
B_r[A]:=\{x\in\Omega\,|\,\,{\rm dist\,}(x,A)<r\}\,.
 $$

The inner product of functions $u,v\in L_2(\Omega)$ is
 $$
(u,v):=\int_\Omega u(x)v(x)\,dx
 $$
where $dx$ is the area element. By $\langle a,b\rangle$ we denote
the inner product of the vectors $a,b \in T\Omega_x$ and put
 $$
(a,b):=\int_\Omega \langle a(x),b(x)\rangle\,dx
 $$
for the vector fields $a,b\in\vec L_2(\Omega)$.

Let $\Delta$ be the Beltrami-Laplace operator, $\nabla$ the
gradient in $\Omega$. The divergence is defined by
 $$
({\rm div\,} a,\varphi)=-\,(a,\nabla \varphi),\qquad \varphi\in
C^\infty_0(\Omega)\,.
 $$
The relation $\Delta={\rm div\,}\nabla$ holds. In some places,
emphasizing the correspondence to the given metric, we write
$\Delta_g,\,\nabla_g,\,{\rm div}_g$, and so on.

Unless otherwise specified, we deal with {\it real} functions and
fields. However, later on the $\mathbb C$-valued functions are
also in the use.
\smallskip

\noindent$\bullet$\,\,\,A function $u$ obeying $\Delta u=0$ is
{\it harmonic}. Harmonic functions are smooth in ${\rm
int\,}\Omega$.

Let $\omega\subset\Omega$. Two smooth functions $u$ and $v$ are
called {\it conjugate} (we write $u\overset{\omega}\sim v$), if
$\langle \nabla u,\nabla v\rangle=0$ and $|\nabla u|=|\nabla v|$
holds everywhere in $\omega$. Note that $u\overset{\omega}\sim v$
and $u\overset{\omega}\sim \pm\, v+\rm const$ are equivalent. The
fields $\nabla u$ and $\nabla v$ are also called conjugate, and we
write $\nabla u\overset{\omega}\sim\nabla v$. We also agree in the
case $\omega={\rm int\,}\Omega$ to write just $u\sim v$ and
$\nabla u\sim\nabla v$.

As is well known, if $\Omega$ is orientable then it supports the
pairs of conjugate functions. The orientation can be fixed by the
choice of the continuous family of isometries ({\it rotations})
$\{\Phi(x)\in {\rm End\,}T\Omega_x\,|\,\,x\in\omega\}$ such that
 \begin{equation}\label{Eq rotation Phi}
\Phi^*=\Phi^{-1},\,\,\Phi^2=-I,\quad\text{and}\quad \nabla
v=\Phi\nabla u \quad \text{in} \,\,\Omega
 \end{equation}
for a pair $u,v$ provided ${\rm const}\not\equiv u\sim v$. The
third relation is the Cauchy-Riemann conditions on $u$ and $v$. By
(\ref{Eq rotation Phi}) one has $\Delta v={\rm div\,}\Phi\nabla
u\equiv 0$ and $\Delta u=-{\rm div\,}\Phi\nabla v\equiv 0$, so
that conjugacy implies harmonicity.
\smallskip

\noindent{$\bullet$}\,\,\,Also, the local congugacy implies
orientability.
 \begin{Lemma}\label{L conjugacy at Gamma}
Let the functions $u\not\equiv{\rm const}$ and $v$ be harmonic in
$\Omega$ and $u\overset{\Gamma'}\sim v$ hold for a segment
$\Gamma'\subset\Gamma$ of positive length. Then $u\sim v$ in
$\Omega$, whereas $\Omega$ is orientable.
 \end{Lemma}
\begin{proof}
$\bf 1.$\,\,\,Let $\omega$ be a (small enough) neighborhood of
$\Gamma'$ diffeomorphic to a disk in ${\mathbb R}^2$ and such that
$\partial\omega\supset\Gamma'$. By the Poincare Theorem, there is
a harmonic $v'$ provided $v'\overset{\omega}\sim u$ and, in
particular, $v'\overset{\Gamma'}\sim u$. By the uniqueness of
harmonic continuation, the latter implies $v'=\pm v+c$, because
$v'$ and $v$ have the same Cauchy data at $\Gamma'$ for the
properly chosen sign and constant. Hence, we have
$u\overset{\omega}\sim v$.

Fix the points $x\in{\rm int\,}\Omega$ and $x'\in\omega$. Choose a
smooth simple curve $l$ connecting $x$ and $x'$ and its (small
enough) neighborhood $\omega'$ diffeomorphic to a disk.  Let $v''$
be conjugate to $u$ in $\omega'$. Since the conjugate function is
determined uniquely up to the sign and constant summand, one can
take $v''=v'=v$ in $\omega\cap\omega'$ that obviously leads to
$v\overset{\omega'}\sim u$. Since $x$ is arbitrary, we conclude
that $v\sim u$ in $\Omega$. Thus, $u\overset{\Gamma'}\sim v$ leads
to $u\sim v$ and $\nabla u\sim \nabla v$.
\smallskip

$\bf 2.$\,\,\,The conjugated fields $\nabla u$ and $\nabla v$ may
have (the same) zeros $x_1, x_2, \dots$ in $\omega$. By
harmonicity, these zeros are isolated in ${\rm int\,}\Omega$,
whereas the open set $\Omega_0:={\rm int\,}\Omega\setminus\{x_1,
x_2, \dots\}$ is a chart with the coordinates $u,v$ oriented by
the Cauchy-Riemann conditions (\ref{Eq rotation Phi}).

Choose a sequence of (small enough) positive $r_1, r_2,\dots$ such
that each $B_{r_j}[x_j]\subset \Omega$ is diffeomorphic to a disk,
and endow it with the orientation consistent with the orientation
of $\Omega_0$. Thus, ${\rm int\,}\Omega$ admits the oriented atlas
$\{\Omega_0, B_{r_1}[x_1],B_{r_2}[x_2],\dots\}$, i.e., is
orientable. Hence, $u\overset{\Gamma'}\sim v$ yields orientability
of $\Omega$.
\end{proof}

\subsubsection*{Harmonic fields}
\noindent$\bullet$\,\,\, A vector field $h$ is said to be {\it
harmonic} in $\Omega$ if for every $x\in{\rm int\,}\Omega$ there
is a disk $D:=B_r[x]$ oriented by a family of rotations $\Phi_D$
such that
 $$
{\rm div\,} h = {\rm div\,}\Phi_D h= 0\qquad {\rm in}\,\,\,D\,.
 $$
By the Poincare Theorem, harmonic fields are locally {\it
potential}: there is a harmonic function $u$ such that $h=\nabla
u$ in $D$. The potential $u$ has the local conjugate $v$, so that
$u\overset{D}\sim v$ and $h=-\Phi_D\nabla v$ holds.

Let
 $$
{\mathscr H}:=\{h\in\vec L_2(\Omega)\,|\,\,h\,\,\text{is harmonic
in}\,\,\Omega\},\quad {\mathscr E}:=\{h\in{\mathscr
H}\,|\,\,h=\nabla u\,\,\text{in}\,\,\Omega\}
 $$
be the space of harmonic fields and its subspace of potential
harmonic fields. In the second definition, the potential $u$
belongs to the Sobolev class $H^1(\Omega)$. The smooth fields are
dense in ${\mathscr H}$ and ${\mathscr E}$.
\smallskip

\noindent$\bullet$\,\,\,Depending on the topology of $\Omega$, the
subspace ${\mathscr N}:={\mathscr H}\ominus{\mathscr E}$ may be
nontrivial. In the latter case, it consists of the so-called
Neumann fields, which are tangent on $\Gamma$. Indeed, for a
smooth $f$ in (\ref{Eq Dir data}) and $n\in{\mathscr N}$ one has
$\nabla u^f\in{\mathscr E}$ and
 $$
0=(\nabla u^f,n)=\int_\Omega \langle\nabla
u^f,n\rangle\,dx=\int_\Gamma f\langle n,\nu\rangle\,ds-\int_\Omega
u^f\,{\rm div\,} n\,dx=\int_\Gamma f\langle n,\nu\rangle\,ds
 $$
($ds$ is the length element of the metric $g$ on $\Gamma$), which
leads to $\langle n,\nu\rangle=0$ on $\Gamma$ by arbitrariness of
$f$. Also, ${\rm dim\,}{\mathscr N}$ is finite and determined by
topology of $\Omega$: see, e.g., \cite{Sch}.

Let $\Omega$ be orientable and (globally) oriented by a family of
rotations $\Phi$. The family determines the unitary operator in
${\mathscr H}$ that acts point-wise by $(\Phi h)(x)=\Phi(x)h(x)$;
we denote it by the same symbol $\Phi$. The subspace
 $$
{\mathscr E}^{\rm c}\,:=\,{\mathscr E}\cap\Phi{\mathscr E}
 $$
is invariant with respect to $\Phi$. A field $h\in{\mathscr H}$
belongs to ${\mathscr E}^{\rm c}$ iff it has the conjugate $h^{\rm
c}\sim h$ in $\Omega$. In such a case, the relations $h=\nabla u$,
$h^{\rm c}=\nabla v=\Phi \nabla u$ (or $h^{\rm c}=\nabla v=-\Phi
\nabla u$), $u\sim v$ hold. As is well known, for any orientable
$\Omega$ the subspace ${\mathscr E}^{\rm c}$ is nontrivial and,
moreover, ${\rm dim\,}{\mathscr E}^{\rm c}=\infty$ holds.

\subsubsection*{Criterion of orientability}

\noindent$\bullet$\,\,\,Denote $\dot L_2(\Gamma):=\{f\in
L_2(\Gamma)\,|\,\,\int_\Gamma f\,ds=0\}$ and recall that
$\Gamma=\Gamma_1\cup\dots\cup\Gamma_N$. For a smooth $f$, the
Green formula implies
 $$
0=\int_\Omega \Delta u^f\,dx=\int_\Gamma\partial_\nu
u^f\,ds=\int_\Gamma \Lambda f\,ds=\int_{\Gamma_1} \Lambda
f\,ds+\dots +\int_{\Gamma_N} \Lambda f\,ds,
 $$
so that $\Lambda f\in \dot L_2(\Gamma)$. Moreover, the following
is valid.
\smallskip

Recall that $\Lambda$ is a positive selfadjoint 1-st order
pseudo-differential operator in $L_2(\Gamma)$, ${\rm
Dom\,}\Lambda=H^1(\Gamma)$, whereas ${\rm Ker\,}\Lambda=\{{\rm
const}\}$ and ${\rm Ran\,}\Lambda=\dot L_2(\Gamma)$ holds. Note in
addition that the length element $ds$ (the metric on $\Gamma$) is
determined by the principal symbol of $\Lambda$ \cite{Taylor}.
 \begin{Lemma}\label{L about psi j}
Let $\Omega$ be orientable and $N\geqslant 2$. Let a smooth $f$ be
such that $\nabla u^f\in{\mathscr E}^{\rm c}$. Then
 \begin{equation}\label{Eq Lambda f in dot L j}
(\Lambda f)\big|_{\Gamma_j}\in \dot L_2(\Gamma_j),\qquad
j=1,\dots,N
 \end{equation}
holds.
 \end{Lemma}
\begin{proof}
The harmonic potential fields
 $$
d_j:=\nabla u^{\psi_j},\qquad
\psi_j\big|_{\Gamma_k}=\delta_{jk},\qquad j,k=1,\dots,N
 $$
are normal to $\Gamma$ and, hence, $\Phi d_j$ are tangent.
Therefore $\Phi d_j$ is a harmonic field tangent to $\Gamma$,
i.e., $\Phi d_j\in{\mathscr N}={\mathscr H}\ominus{\mathscr E}$.

Since $\nabla u^f\in{\mathscr E}^{\rm c}$, one has $\Phi \nabla
u^f\in{\mathscr E}^{\rm c}$, so that $\Phi\nabla u^f\bot{\mathscr
N}$. The latter implies
 \begin{align*}
& 0=(\Phi\nabla u^f,\Phi d_j)=(\nabla u^f,d_j)=\int_\Gamma
f\langle d_j,\nu\rangle\,ds=\int_\Gamma f\langle
\nabla u^{\psi_j},\nu\rangle\,ds=\\
& =\int_\Gamma f\Lambda \psi_j\,ds=\int_\Gamma \Lambda
f\,\psi_j\,ds=\int_{\Gamma_j}\Lambda f\,ds
 \end{align*}
and we arrive at (\ref{Eq Lambda f in dot L j}).
\end{proof}
\smallskip

\noindent$\bullet$\,\,\,Let $\gamma$ be a continuous tangent field
of unit vectors on $\Gamma$. For functions on the boundary, by
$\dot f:=\partial_\gamma f$ we denote the derivative with respect
to the length $s$  in direction $\gamma$, so that $\nabla_\Gamma
f= \dot f\gamma$. Also, note the evident relation $\dot
f\big|_{\Gamma_j}\in \dot L_2(\Gamma_j)$. For the solution to
(\ref{Eq Laplace}) and (\ref{Eq Dir data}) one has
 $$
\nabla u^f=\nabla_\Gamma f + (\partial_\nu u^f)\nu=\dot f \gamma+
(\Lambda f)\,\nu \qquad {\rm on}\,\,\,\Gamma.
 $$

Assume that $\Omega$ is oriented by $\Phi$ and $\Phi \nu=\gamma$.
Let the smooth $f$ and $p$ be such that ${\rm const}\not\equiv
u^f\sim u^p$ and $\nabla u^p=\Phi\nabla u^f$. Then the relations
 \begin{align*}
& \Lambda p=\langle \nu,\nabla u^p\rangle=\langle \nu,\Phi\nabla
u^f\rangle=-\langle
\Phi\nu,\nabla u^f\rangle =-\langle\gamma,\nabla u^f\rangle=-\dot f,\\
& \Lambda f=\langle \nu,\nabla u^f\rangle=-\langle \nu,\Phi\nabla
u^p\rangle=\langle \Phi\nu,\nabla u^p\rangle =\langle\gamma,\nabla
u^p\rangle=\dot p
 \end{align*}
hold and lead to a `Hamiltonian' system of the form
 \begin{equation}\label{Eq Hamilton}
\dot f=-\Lambda p,\quad \dot p=\Lambda f\qquad {\rm
on}\,\,\,\Gamma.
 \end{equation}
In addition, note that these relations are consistent with
(\ref{Eq Lambda f in dot L j}) on each $\Gamma_j$.
\smallskip

\noindent$\bullet$\,\,\,Assume that the smooth functions $f$ and
$p$ are such that (\ref{Eq Hamilton}) is satisfied {\it at least
on one} component $\Gamma_j$ of the boundary. Then $u^f\sim u^p$
holds, whereas $\Omega$ is orientable.

Indeed, we have
 \begin{align*}
& \langle\nabla u^f,\nabla u^p\rangle=\langle\dot f \gamma+
(\Lambda f)\nu\,,\,\dot p \gamma+ (\Lambda p)\nu\rangle =\dot
f\dot p+\Lambda
f\,\Lambda p\,\overset{{\rm see\,}(\ref{Eq Hamilton})}=\,0,\\
& |\nabla u^f|^2=\dot f^2+ (\Lambda f)^2\,\overset{{\rm
see\,}(\ref{Eq Hamilton})}=\,(\Lambda p)^2+\dot p^2 =|\nabla
u^p|^2 \qquad {\rm on}\,\,\,\Gamma_j,
 \end{align*}
so that $u^f\overset{\Gamma_j}\sim u^p$. The latter, by the Lemma
\ref{L conjugacy at Gamma}, implies $u^f\overset{\Omega}\sim u^p$
and follows to to orientability of $\Omega$.
\smallskip

Summarizing, we arrive at the following criterion.
 \begin{Theorem}\label{Th Criterion}
The manifold $\Omega$ is orientable if and only if there are a
tangent field $\gamma$ and a pair of smooth functions
$f\not\equiv{\rm const}$ and $p$ on $\Gamma$ such that (\ref{Eq
Hamilton}) holds at least on one of the components $\Gamma_j$ of
the boundary.
 \end{Theorem}
By Lemma \ref{L conjugacy at Gamma}, this statement remains true
if we replace `one of the components $\Gamma_j$' with `any segment
$\Gamma'$ on one of the components $\Gamma_j$'.
\smallskip

\noindent$\bullet$\,\,\,Let $\Gamma$ consist of the single
component. Introduce the integration $J:\dot L_2(\Gamma)\to\dot
L_2(\Gamma)$ by $\partial_\gamma J={\rm id}$. In this case, one
has
 $$
\dot f=-\Lambda p=-\Lambda J\dot p=-(\Lambda J)^2\dot f
 $$
that leads to $[I\!+\!(\Lambda J)^2]\dot f=0$ and $\dot p=\Lambda
J\dot f$. In the case of the orientable $\Omega$, the latter
relations enable one to find the traces of the conjugated
functions $u^f$ and $u^p$ at the boundary via $\Lambda$: see
\cite{BCald}. The above established criterion can be formulated as
follows.
 \begin{Corollary}\label{Cor 1}
Let $\Gamma$ consist of a single component. Then $\Omega$ is
orientable iff \,${\rm Ker\,}[I\!+\!(\Lambda J)^2]\not=\{0\}$.
 \end{Corollary}

So, given the DN-map, one can determine whether the manifold is
orientable or not.

\section{Moebius band}

\subsubsection*{Attributes}
\noindent$\bullet$\,\,\,Let $\Gamma$ be diffeomorphic to a circle
in ${\mathbb R}^2$, $ds$ the length element on $\Gamma$. For
$m,m'\in\Gamma$, we put $m':=-m$ if ${\rm
dist}_\Gamma(m,m')=\frac{{\rm mes\,}\Gamma}{2}$ holds.

Let ${\mathbb M}:=\Gamma\times[-1,1]$ be the cylinder. For
$x=\{m,\alpha\}\in{\mathbb M}$ we put $-x:=\{-m,-\alpha\}$ and
denote by $\tau$ the involution $\tau: x\mapsto -x$. The relation
$x\sim x' \Leftrightarrow\tau(x)=x'$ is an equivalence on
${\mathbb M}$.

The Moebius band is $M:={\mathbb M}/\tau$, so that ${\mathbb M}$
is the double covering of $M$. By $\pi:{\mathbb M}\to M$ we denote
the natural projection, which is a local diffeomorphism. The
boundary $\partial{\mathbb M}$ consists of two components
$\Gamma_\pm=\{x=\{m,\pm 1\}\,|\,\,m\in\Gamma\}$. The boundary
$\partial M=\pi(\partial {\mathbb M})=\pi(\Gamma_+)=\pi(\Gamma_-)$
is identified with $\Gamma$ by $\pi(\{m,\pm1\})\equiv m$.
\smallskip

Let $M$ be endowed with a metric $g$, $ds$ be the length element
of $g$ on $\Gamma$. By $\Delta_g,\,\nabla_g,\,\Lambda_g,\dots$ we
denote the corresponding operations in $M$.

The metric on $M$ induces the metric ${\rm g\,}=\pi_*g$ on
${\mathbb M}$; recall that ${\rm g\,}(a,b)\big|_x:=g(D_\pi a,D_\pi
b)\big|_{\pi(x)}$ for the tangent vectors $a,b\in T{\mathbb M}_x$,
where $D_\pi$ is the differential of the projection. Also, $\pi$
is a local isometry, i.e., ${\rm dist}_{\mathbb M}(x,y)={\rm
dist}_M(\pi(x),\pi(y))$ holds for the close enough $x$ and $y$. As
is easy to see, the induced metric obeys
 \begin{equation}\label{Eq tau-inv of g}
\tau_*{\rm g\,}\,=\,{\rm g\,}\,.
 \end{equation}
Simplifying the notation, we denote $\Delta:=\Delta_{\rm
g},\,\nabla:=\nabla_{\rm g},\,\Lambda:=\Lambda_{\rm g\,}$ and so
on.
\smallskip

In contrast to $M$, its covering ${\mathbb M}$ is orientable, and
in the subsequent we assume ${\mathbb M}$ to be oriented by a
rotation $\Phi$. Its boundary is also oriented by the tangent
field $\gamma=\Phi\nu$, where $\nu$ is the outward normal to
$\partial{\mathbb M}$. There are two orientations of the boundary
$\Gamma=\partial M$. For definiteness, we put it to be oriented by
the tangent field $D_\pi[\gamma\big|_{\Gamma_+}]$, and denote this
field by the same $\gamma$.
\smallskip

\noindent$\bullet$\,\,\,A function ${\rm u}$ on ${\mathbb M}$ is
said to be even (odd) if ${\rm u}={\rm u}\circ\tau$ (${\rm
u}=-{\rm u}\circ\tau$) holds. If ${\rm u}$ is even, there is a
function $u$ on $M$ such that ${\rm u} =u\circ\pi$. If ${\rm f}$
is even (odd), then $\dot {\rm f}:=\partial_\gamma{\rm f}$ is odd
(even), and the following relations can be easily derived and will
be used later:
 \begin{equation}\label{Eq auxill}
(f\circ\pi)^2=f^2\circ\pi;\quad\partial_\gamma(f\circ\pi)=\sigma\,(\partial_\gamma
f)\circ\pi \qquad {\rm on}\,\,\partial {\mathbb M}\,,
 \end{equation}
where $f$ is a function on $\Gamma$ and
$\sigma\big|_{\Gamma_\pm}:=\pm 1$.

As is easy to verify, the relation
 \begin{equation*}
\Delta(u\circ \pi)\,=\,(\Delta_g u)\circ \pi\qquad {\rm
in}\,\,\,{\rm int}\,\,{\mathbb M}
 \end{equation*}
holds and $\Delta$ preserves the parity. As a consequence, if
${\rm u}={\rm u}^{\rm f}(x)$ satisfies
 \begin{equation}\label{Eq Dir prob on mbM}
\Delta{\rm u}\,=\,0\quad {\rm in}\,\,\,{\rm int}\,{\mathbb
M},\quad {\rm u}={\rm f}\quad {\rm on}\,\,\partial {\mathbb M}
 \end{equation}
and ${\rm f}=f\circ\pi$ then one has the relations
 \begin{equation}\label{Eq Lambda pi=pi Lambda}
{\rm u}^{f\circ\pi}=u^f\circ \pi, \quad
\Lambda(f\circ\pi)=(\Lambda_g f)\circ\pi\qquad {\rm
on}\,\,\partial{\mathbb M}\,,
 \end{equation}
where $u^f$ solves (\ref{Eq Laplace}),(\ref{Eq Dir data}) on $M$.
So, $\Lambda$ also preserves the parity.
\smallskip

\noindent$\bullet$\,\,\,The plan of solving the EIT problem for
$M$ is, loosely speaking, as follows. First, we show that
$\Lambda_g$ determines (up to isometry) the analytic function
algebra on the cylinder ${\mathbb M}$, which does exist owing to
its orientability. Then, by the use of the technique \cite{BCald},
we construct a homeomorphic copy ${\mathbb M}^{\,'}$ of ${\mathbb
M}$ and endow it with a relevant metric ${\rm g}'$, which obeys
(\ref{Eq tau-inv of g}). At last, we determine a copy $M'$ of $M$
and supply it with the metric $g'=\pi^{-1}_*{\rm g}'$. As a
result, the manifold $(M',g')$ turns out to be isometric to the
(unknown) original $(M,g)$ and, thus, provides the solution of the
problem.

\subsubsection*{Harmonicity in $({\mathbb M},{\rm g\,})$}
\noindent$\bullet$\,\,\,Let $\phi:={\rm u}^{\rm f}$ be the
solution of (\ref{Eq Dir prob on mbM}) for ${\rm f} =\pm 1$ on
$\Gamma_\pm$. The harmonic potential field $\nabla
\phi\in{\mathscr E}$ in ${\mathbb M}$ is normal on $\Gamma_\pm$.
For harmonic fields in ${\mathbb M}$, we have ${\mathscr
H}={\mathscr E}\oplus{\mathscr N}$ and, as is known, ${\rm
dim\,}{\mathscr N}=1$ and ${\mathscr
N}=\{c\,\Phi\nabla\phi\,|\,\,c=\rm const\}$ holds.
 \begin{Lemma}\label{L auxill}
For any smooth ${\rm f}$, there is a smooth ${\rm p}$ the a
constant $c$ such that the equality
 \begin{equation}\label{Eq 2.5}
\Phi\nabla {\rm u}^{\rm f}=\nabla{\rm u}^{\rm p}+c\,\Phi\nabla\phi
 \end{equation}
holds, where
 \begin{equation}\label{Eq 2.6}
\dot{\rm p}=\Lambda{\rm f}-c\,\Lambda\phi\quad{\rm
on\,\,}\partial{\mathbb M},\qquad
c=\frac{\int_{\Gamma_+}\Lambda{\rm
f}\,ds-\int_{\Gamma_-}\Lambda{\rm f}\,ds}{\|\nabla\phi\|^2}\,.
 \end{equation}
 \end{Lemma}
\begin{proof}
The equality (\ref{Eq 2.5}) follows from $\Phi\nabla {\rm u}^{\rm
f}\in{\mathscr H}$ and ${\mathscr H}={\mathscr E}\oplus{\mathscr
N}$. Multiplying it by $\gamma$, one has
 \begin{align*}
&\langle\gamma,\Phi\nabla{\rm u}^{\rm
f}\rangle=-\langle\Phi\gamma,\nabla{\rm u}^{\rm
f}\rangle=\langle\nu,{\rm u}^{\rm f}\rangle=\Lambda{\rm
f}\overset{(\ref{Eq
2.5})}=\langle\gamma,\nabla{\rm u}^{\rm p}\rangle+c\,\langle\gamma,\Phi\nabla\Phi\rangle=\\
& = \dot{\rm p}+c\,\langle\nu,\nabla\Phi\rangle=\dot{\rm
p}+c\,\Lambda\phi\,,
 \end{align*}
so that $\dot{\rm p}=\Lambda{\rm f}-c\Lambda\phi$ does hold.
Multiplying by $\Phi\nabla\phi$ and integrating over ${\mathbb
M}$, one gets
 \begin{align*}
&(\Phi\nabla\phi,\Phi\nabla{\rm u}^{\rm f})=(\nabla\phi,\nabla{\rm u}^{\rm f})=
\int_{\partial{\mathbb M}}{\rm f}\,\Lambda\phi\,ds=\int_{\partial{\mathbb M}}\Lambda{\rm f}\,\phi\,ds=\\
&=\int_{\Gamma_+}\Lambda{\rm f}\,ds-\int_{\Gamma_-}\Lambda{\rm
f}\,ds\overset{(\ref{Eq 2.5})}=(\Phi\nabla\phi,\nabla{\rm u}^{\rm
p})+c\,\|\nabla\phi\|^2=c\,\|\nabla\phi\|^2
 \end{align*}
since $\Phi\nabla\phi\in{\mathscr N}$, whereas $\nabla{\rm u}^{\rm
p}\bot{\mathscr N}$. Thus, (\ref{Eq 2.6}) is valid.
\end{proof}
\smallskip

\noindent$\bullet$\,\,\,As a consequence of (\ref{Eq 2.5}), we
have the following.
 \begin{Corollary}\label{Cor 2}
The relations
 \begin{equation}\label{Eq with odd p}
\Phi\nabla{\rm u}^{f\circ\pi}=\nabla {\rm u}^{\rm p}\in{\mathscr
E},\quad\dot{\rm p}=(\Lambda_g f)\circ\pi,\quad {\rm
p}\circ\tau=-{\rm p}\qquad {\rm in}\,\,{\mathbb M}
 \end{equation}
hold for any $f$ smooth on $\Gamma=\partial M$.
 \end{Corollary}
Indeed, the function ${\rm f}=f\circ\pi$ is even on ${\mathbb M}$.
Hence, by (\ref{Eq Lambda pi=pi Lambda}) the function $\Lambda{\rm
f}$ is also even and, as a result, we have $c=0$  in (\ref{Eq
2.6}). In the meanwhile, the function ${\rm p}$ satisfies $\dot
{\rm p}=\Lambda{\rm f}\overset{(\ref{Eq Lambda pi=pi
Lambda})}=(\Lambda_g f)\circ\pi$ and, hence, one can choose it to
be odd. In what follows we accept such a choice.

Let $J:\dot L_2(\Gamma)\to \dot L_2(\Gamma),\,\,\partial_\gamma
J=\rm id$ be the corresponding integration. Then, in addition to
(\ref{Eq with odd p}) one has
  \begin{equation}\label{Eq for p}
{\rm p}\,=\,\sigma\left[(J\Lambda_g
f)\circ\pi+b\,\right]=\sigma\left[(J\Lambda_g
f)\circ\pi\right]+b\phi\quad {\rm on}\,\,\partial{\mathbb M}\,,
 \end{equation}
where $\sigma\big|_{\Gamma_\pm}=\pm 1$ and $b$ is a constant.
Respectively, one gets
 \begin{equation}\label{Eq *}
\nabla {\rm u}^{\rm p}=\nabla{\rm u}^{\sigma\left[(J\Lambda_g
f)\circ\pi\right]}+b\nabla\phi\qquad {\rm in}\,\,{\mathbb M}.
 \end{equation}
To find $b$ we use the orthogonality $\nabla{\rm
u}^{f\circ\pi}\bot\Phi\nabla\phi$ in ${\mathscr H}$: the relations
 \begin{align*}
& 0=-(\nabla{\rm u}^{f\circ\pi},\Phi\nabla\phi)=(\Phi\nabla{\rm
u}^{f\circ\pi},\nabla\phi)\overset{(\ref{Eq with odd p})}=(\nabla
{\rm u}^{\rm p},\nabla\phi)\overset{(\ref{Eq *})}=(\nabla{\rm
u}^{\sigma\left[(J\Lambda_g
f)\circ\pi\right]},\nabla\phi)+\\
& +b\,(\nabla\phi,\nabla\phi)\overset{\rm
int.\,\,by\,\,parts}=\int_{\partial{\mathbb
M}}\sigma\left[(J\Lambda_g
f)\circ\pi\right]\langle\nu,\nabla\phi\rangle\,ds+b\int_{\partial{\mathbb M}}\phi\langle\nu,\nabla\phi\rangle\,ds=\\
& =2\int_{\Gamma_+}\left[(J\Lambda_g
f)\circ\pi\right]\Lambda\phi\,ds+2b\int_{\Gamma_+}\Lambda\phi\,ds
 \end{align*}
hold and imply
 \begin{equation}\label{Eq for b}
b=-\frac{\int_{\Gamma_+}\left[(J\Lambda_g
f)\circ\pi\right]\Lambda\phi\,ds}{\int_{\Gamma_+}\Lambda\phi\,ds}\,.
 \end{equation}
\smallskip

\noindent$\bullet$\,\,\,The first relation in (\ref{Eq with odd
p}) shows that ${\rm u}^{f\circ\pi}$ and ${\rm u}^{\rm p}$ are
conjugate by Cauchy-Riemann and, hence, the $\mathbb C$-valued
function ${\rm w}={\rm u}^{f\circ\pi}+i{\rm u}^{\rm p}$ is
holomorphic in ${\mathbb M}$. Its boundary value ({\it trace})
${\rm Tr\,}{\rm w}:=w\big|_{\partial{\mathbb M}}$ is represented
by (\ref{Eq for p}) and (\ref{Eq for b}):
 \begin{equation*}
{\rm Tr\,}w={\rm Tr\,}{\rm u}^{f\circ\pi}+{\rm Tr\,}{\rm u}^{\rm
p}={f\circ\pi}+i{\rm p}={f\circ\pi}+i\sigma\left[(J\Lambda_g
f)\circ\pi+b\right]\,.
 \end{equation*}
However, there is a disadvantage of this representation. When
solving the EIT problem, the observer possesses the operator
$\Lambda_g$ but not $\Lambda$, which enters in (\ref{Eq for b}).
To eliminate it, we use the following artificial trick.

To simplify the notation, denote $q:=J\Lambda_g f$. The function
${\rm w}^2$ is holomorphic in ${\mathbb M}$ and, hence, $\Re{\rm
w}^2=({\rm u}^{f\circ\pi})^2-({\rm u}^{\rm p})^2$ is harmonic in
${\mathbb M}$. Therefore, for ${\rm v}=({\rm
u}^{f\circ\pi})^2-({\rm u}^{\rm p})^2$ we have
 \begin{align}
\notag & \partial_\nu{\rm v}=\Lambda({\rm
v}\big|_{\partial{\mathbb M}})=\Lambda[(f\circ\pi)^2-{\rm
p}^2]\overset{(\ref{Eq auxill}),(\ref{Eq Lambda pi=pi
Lambda}),(\ref{Eq for p})}=
(\Lambda_g f^2)\circ\pi-\Lambda[\sigma q\circ\pi+b\phi]^2=\\
\notag & =(\Lambda_g f^2)\circ\pi-\Lambda[(q\circ\pi)^2+2b(q\circ\pi)\sigma\phi+b^2\phi^2]=\\
\label{Eq d nu v 1} &  =(\Lambda_g f^2)\circ\pi-(\Lambda_g
q^2)\circ\pi-2b\Lambda(q\circ\pi)=[\Lambda_g f^2-\Lambda_g
q^2-2b\Lambda_g q]\circ\pi\,,
 \end{align}
where $\phi^2=\sigma\phi=1$ and $\Lambda b^2=b^2\Lambda 1=0$ were
used. On the other hand, we have
 \begin{align}
\notag & \partial_\nu{\rm v}=\partial_\nu({\rm u}^{f\circ\pi})^2-\partial_\nu({\rm u}^{\rm p})^2=
2{\rm u}^{f\circ\pi}\partial_\nu{\rm u}^{f\circ\pi}-2{\rm u}^{\rm p}\partial_\nu{\rm u}^{\rm p}=\\
\label{Eq d nu v 2} & = 2(f\circ\pi)\Lambda(f\circ\pi)-2{\rm
p}\Lambda{\rm p}=2(f\Lambda_g f)\circ\pi-2{\rm p}\Lambda{\rm p}\,.
 \end{align}
In the meanwhile, multiplying the first relation in (\ref{Eq with
odd p}) by $\nu$ at $\partial{\mathbb M}$, we have
 \begin{align*}
& \Lambda{\rm p}=\langle\nu,\nabla{\rm u}^{\rm p}\rangle=\langle\nu,\Phi\nabla{\rm u}^{f\circ\pi}\rangle=
-\langle\Phi\nu,\nabla{\rm u}^{f\circ\pi}\rangle=-\langle\gamma,\nabla{\rm u}^{f\circ\pi}\rangle=\\
& =-\,\partial_\gamma(f\circ\pi)\overset{(\ref{Eq
auxill})}=-\,\sigma\,\partial_\gamma f\circ\pi,
 \end{align*}
which follows to
 \begin{align*}
& {\rm p}\Lambda{\rm p}=-{\rm p}\,\sigma\,(\partial_\gamma
f)\circ\pi=-[\sigma q\circ\pi+b\phi]\,\sigma\,(\partial_\gamma
f)\circ\pi=-[q\circ\pi]\,[(\partial_\gamma f)\circ\pi]-\\
& -b\,[(\partial_\gamma f)\circ\pi]=-[q\partial_\gamma
f+b\,\partial_\gamma f]\circ\pi\,.
 \end{align*}
Substituting to (\ref{Eq d nu v 2}), we get
 \begin{equation}\label{Eq d nu v 3}
\partial_\nu{\rm v}=2\,[f\Lambda_g f+q\partial_\gamma f+b\,\partial_\gamma
f]\circ\pi\,.
 \end{equation}
At last, equating the results in (\ref{Eq d nu v 3}) and (\ref{Eq
d nu v 1}) one easily arrives at
 \begin{equation}\label{Eq for b final}
b=\frac{\frac{1}{2}\,[\Lambda_g f^2-\Lambda_g
q^2]-f\Lambda_gf-q\dot f}{\dot f+\Lambda_g
q},\quad\text{where}\quad q=J\Lambda_g f,\,\,\dot
f=\partial_\gamma f\,.
 \end{equation}
The remarkable feature of this representation is that, first, it
contains $\Lambda_g$ only (does not contain $\Lambda$) and,
second, the terms entering in the right hand side are the
functions of $x\in\Gamma$ but the ratio is constant. Also, the
denominator
 $$
\dot f+\Lambda_g q=\dot f+\Lambda_g J\Lambda_gf=\dot f+\Lambda_g
J\Lambda_gJ\dot f=[I+(\Lambda_g J)^2]\dot f
 $$
can not have too many zeros on $\Gamma$ since, by Corollary
\ref{Cor 1}, one has ${\rm Ker\,}[I+(\Lambda_g J)^2]=\{0\}$ for
the nonorientable $M$.

\subsubsection*{Algebra ${\mathfrak A}({\mathbb M})$}
\noindent$\bullet$\,\,\,Let
 $$
{\mathfrak A}({\mathbb M}):=\left\{{\rm w}={\rm u}+i{\rm
v}\,|\,\,{\rm u},{\rm v}\in C({\mathbb M}),\,\,\nabla{\rm
v}=\Phi\nabla{\rm u}\,\,\,{\rm in}\,\,{\rm int\,}{\mathbb
M}\right\}
 $$
be the Banach algebra of holomorphic continuous functions with the
norm $\|{\rm w}\|={\rm sup}_{\mathbb M} |{\rm w}|$. Its smooth
elements ${\mathfrak A}^\infty({\mathbb M}):={\mathfrak
A}({\mathbb M})\cap C^\infty({\mathbb M};\mathbb C)$ are dense in
${\mathfrak A}({\mathbb M})$. A special feature of the algebra
${\mathfrak A}({\mathbb M})$ is the presence of the involution
 $$
{\rm w}\mapsto{\rm w}^*:=\overline{{\rm w}\circ\tau}\,.
 $$
By
 $$
{\mathfrak A}_*({\mathbb M}):=\{{\rm v}\in{\mathfrak A}({\mathbb
M})\,|\,\,{\rm w}^*={\rm w} \},\quad{\mathfrak
A}^\infty_*({\mathbb M}):={\mathfrak A}_*({\mathbb M})\cap
C^\infty({\mathbb M};\mathbb C)
 $$
we denote the sets of the Hermitian elements. For any element of
the algebra, one represents
 \begin{equation}\label{Eq w=y+iz}
{\rm w}={\rm y\,}+i{\rm z},\qquad {\rm y}=\frac{{\rm v}+{\rm
v}^*}{2}\,,\,\,\,{\rm z}= \frac{{\rm v}-{\rm v}^*}{2i}
 \end{equation}
with the Hermitian ${\rm y}$ and ${\rm z}$.

In accordance with the maximal principle, one has ${\rm
sup}_{\mathbb M} |{\rm w}|={\rm sup}_{\partial {\mathbb M}} |{\rm
w}|$ and, hence, the map
 $$
{\mathfrak A}({\mathbb M})\ni{\rm w}\overset{\rm Tr\,\,}\mapsto
{\rm w}\big|_{\partial{\mathbb M}}\in C(\partial{\mathbb
M};\mathbb C)
 $$
is an isometry on its image. In the meantime, obviously, $\rm Tr$
is an isomorphism of algebras. The (sub)algebra
 $$
{\rm Tr\,}{\mathfrak A}({\mathbb M})\subset C(\partial{\mathbb
M};\mathbb C)
 $$
contains the dense set ${\rm Tr\,}{\mathfrak A}^\infty({\mathbb
M})$ and is isometrically isomorphic to ${\mathfrak A}({\mathbb
M})$ via the map $\rm Tr$.

The functions $\Re{\rm w}$ and $\Im{\rm w}$ of ${\rm w}\in
{\mathfrak A}({\mathbb M})$ can be used as the local (isothermal)
coordinates consistent with the smooth structure of ${\mathbb M}$.
\smallskip

\noindent$\bullet$\,\,\,Let us show that the algebra ${\mathfrak
A}({\mathbb M})$ is determined by the DN map $\Lambda_g$ of the
Moebius band, which is the key fact for the EIT problem.

Each element of the form ${\rm w}={\rm u}^{f\circ \pi}+i{\rm
u}^{\rm p}$ obeying (\ref{Eq with odd p}) and (\ref{Eq for b
final}), is Hermitian. This is a simple consequence of the fact
that $\Re {\rm w}$ and $\Im {\rm w}$ are even and odd
respectively. It is easy to check that the converse is also valid.
As a result, passing to the traces on $\partial{\mathbb M}$, we
have the representation
 \begin{align}
\notag & {\rm Tr\,}{\mathfrak A}^\infty_*({\mathbb M})\,=\\
\label{Eq repres Hermite elements}
&=\{w=f+i\,\sigma\,[(J\Lambda_gf)+b]\,|\,\,f\in
C^\infty(\Gamma),\,b\,\,\,\text{obeys}\,\,(\ref{Eq for b
final})\}\,.
 \end{align}
By (\ref{Eq w=y+iz}) and (\ref{Eq repres Hermite elements}), for
any ${\rm w}\in{\mathfrak A}^\infty({\mathbb M})$ one has
 \begin{align}
\notag & {\rm Tr\,}{\mathfrak A}^\infty({\mathbb M})\,=\\
\notag &= \{w=y+iz\,|\,\,y=f+i\sigma[J\Lambda_g
f+b],\,z=f'+i\sigma[J\Lambda_gf'+b']\,;\\
\label{Eq final repres smooth Tr A(M)} & b
\,\,\text{and}\,\,b'\,\,\text{obey}\,\,(\ref{Eq for b
final})\,\,\text{for}\,\,f\,\text{and}\,f'\,\,\,\text{respectively}\,\}\,.
 \end{align}

Summarizing and denoting by ${\mathfrak A}\cong{\mathfrak B}$ the
isomorphic isometry of algebras, we arrive at the following scheme
of determination of ${\mathfrak A}({\mathbb M})$ via the DN-map:
 \begin{equation}\label{Eq FINISH}
\Lambda_g\overset{(\ref{Eq final repres smooth Tr
A(M)})}\Rightarrow {\rm Tr\,}{\mathfrak A}^\infty({\mathbb
M})\Rightarrow{\rm clos\,}_{C(\Gamma;\mathbb C)}{\rm
Tr\,}{\mathfrak A}^\infty({\mathbb M})={\rm Tr\,}{\mathfrak
A}({\mathbb M})\cong{\mathfrak A}({\mathbb M})\,.
 \end{equation}
Thus, $\Lambda_g$ determines the algebra ${\mathfrak A}({\mathbb
M})$ up to an isometric isomorphism.

\subsubsection*{Determination of $(M,g)$}
\noindent$\bullet$\,\,\,Recall the well-known notions and facts
(see, e.g., \cite{Gamellin,Royden})

A {\it character} of the complex commutative Banach algebra is a
nonzero homomorphism $\chi:{\mathfrak A}\to{\mathbb C}$. The set
of characters ({\it spectrum} of ${\mathfrak A}$) is denoted by
$\hat{\mathfrak A}$ and endowed with the canonical Gelfand
($*$-weak) topology. The Gelfand transform ${\mathfrak A}\to
C(\hat{\mathfrak A};{\mathbb C})$ maps $a\in{\mathfrak A}$ to the
function $\hat a\in C(\hat{\mathfrak A};{\mathbb C})$ by $\hat
a(\chi):=\chi(a)$.

We write $S\cong T$ if the topological spaces $S$ and $T$ are
homeomorphic. If the algebras are isometrically isomorphic
(${\mathfrak A}\cong{\mathfrak B}$) then their spectra are
homeomorphic: $\hat{\mathfrak A}\cong\hat{\mathfrak B}$.

For a function algebra ${\mathfrak F}\subset C(T;{\mathbb C})$,
the set ${\mathscr D}$ of the Dirac measures $\delta_t: a\mapsto
a(t),\,\,t\in T$ is a subset of $\hat{\mathfrak F}$. This algebra
is called {\it generic} if $\hat{\mathfrak F}={\mathscr D}$ holds,
which is equivalent to $\hat{\mathfrak F}\cong T$. In this case,
the algebra ${\mathfrak F}$ is isometrically isomorphic to
$C(\hat{\mathfrak F};{\mathbb C})$, the isometry being realized by
the Gelfand transform $a\mapsto\hat a$.
\smallskip

The algebra ${\mathfrak A}({\mathbb M})$ is generic
\cite{Royden,Stout}. Therefore, ${\mathfrak A}({\mathbb
M})\cong{\rm Tr\,}{\mathfrak A}({\mathbb M})$ implies ${\mathbb
M}\cong\widehat{{\mathfrak A}({\mathbb M})}\cong\widehat{{\rm
Tr\,}{\mathfrak A}({\mathbb M})}=:{\mathbb M}^{\,'}$. By the
latter and in accordance with the scheme (\ref{Eq FINISH}), the
DN-map $\Lambda_g$ determines the spectrum ${\mathbb M}^{\,'}$.
\smallskip

\noindent$\bullet$\,\,\,Recall that `to solve the EIT problem' is:
given the DN-map $\Lambda_g$ of the Moebius band $(M,g)$, to
provide a manifold $(M',g')$ such that $\partial M' =\partial M$
and $\Lambda_{g'}=\Lambda_g$ holds. Such a {\it copy} $(M',g')$ of
the original $(M,g)$ is considered to be the solution
\cite{BCald,B Sobolev Geom Rings,B UMN}.
\smallskip

The copy is constructed by means of the following procedure. We
describe it briefly, referring the reader to the papers
\cite{BCald,B Sobolev Geom Rings} for details. Note that, starting
the procedure, we have nothing but the operator $\Lambda_g$ on
$\Gamma$. However, we know a priori that $\Lambda_g$ is the DN-map
of some unknown Moebius band.
\smallskip

{\it Step 1.}\,\,\,Take ${\mathbb
M}=\Gamma\times[-1,1],\,\,\partial{\mathbb
M}=\Gamma_+\cup\Gamma_-$ and identify $\Gamma_+\equiv\Gamma$.
Given $\Lambda_g$, determine the algebra ${\rm Tr\,}{\mathfrak
A}({\mathbb M})$ by (\ref{Eq FINISH}) and find its spectrum
${\mathbb M}^{\,'}$. Applying the Gelfand transform
 $$
{\rm Tr\,}{\mathfrak A}({\mathbb M})\ni w\,\mapsto\,\hat w\in
C({\mathbb M}^{\,'};{\mathbb C})\,,
 $$
we get the relevant copies $\hat w$ of the (unknown) holomorphic
functions ${\rm w}$ in ${\mathbb M}$.
\smallskip

{\it Step 2.}\,\,\,Using $\Re\hat{\rm f}$ and $\Im\hat{\rm f}$ in
capacity of the local coordinates on ${\mathbb M}^{\,'}$, we
supply the spectrum with the structure of a smooth 2d manifold.
\smallskip

{\it Step 3.}\,\,\, By Silov \cite{Gamellin,BCald}, the {\it
boundary} $\partial {\mathbb M}^{\,'}$ is identified as the subset
of ${\mathbb M}^{\,'}$, at which the functions $|\hat{w}|$ attain
the maximum. The boundary is disconnected and consists of two
connected components $\Gamma'_\pm$. Also, we identify the boundary
points by
 $$
\partial{\mathbb M}\ni x\equiv\delta\in\partial {\mathbb M}^{\,'}\,\Leftrightarrow\,
\,{w}(x)=\hat{w}(\delta)\quad\text{for all smooth}\,\,w
 $$
and, thus, attach $\partial {\mathbb M}^{\,'}$ to $\partial
{\mathbb M}$.
\smallskip

{\it Step 4.}\,\,\,The involution $\tau'$ in ${\mathbb M}^{\,'}$,
which is the copy of $\tau$ in ${\mathbb M}$, is determined as
follows. Let $w=w^f\in{\rm Tr\,}{\mathfrak A}^\infty_*({\mathbb
M})$ be a function on $\partial{\mathbb M}$ specified by the
conditions in (\ref{Eq repres Hermite elements}), $\hat w^f$ its
Gelfand transform (a function on ${\mathbb M}^{\,'}$). For a
character $\chi\in {\mathbb M}^{\,'}$, we define
$\chi':=\tau'(\chi)$ if
 $$
\hat w^f(\chi')\,=\,\overline{\hat w^f(\chi)}\quad \text{holds for
all}\,\,\,f\in C^\infty(\Gamma)\,.
 $$
As is easy to recognize, such a definition is motivated by the
relation ${\rm w}(\tau(x))=\overline{{\rm w}(x)}$ on $\mathbb M$
for the functions of the form ${\rm w}={\rm u}^{f\circ\pi}+i{\rm
u}^{\rm p}$ with the even $\Re{\rm w}$ and odd $\Im{\rm w}$.
\smallskip

{\it Step 5.}\,\,\,At the moment, there is no metric on the
spectrum ${\mathbb M}^{\,'}$. In the meantime, it supports the
reserve of functions $\Re\hat w^f\,\Im\hat w^f$, which are the
relevant copies of the (unknown) harmonic functions ${\rm
u}^{f\circ\pi},\,{\rm u}^{\rm p}$ on ${\mathbb M}$. As is known,
this reserve determines a metric $\tilde{\rm g}$ on ${\mathbb
M}^{\,'}$, which provides $\Delta_{\tilde{\rm g}}\Re\hat
w^f=\Delta_{\tilde{\rm g}}\Im\hat w^f=0$, such a metric being
determined up to a conformal deformation. In \cite{LUEns,BCald}
the reader can find concrete tricks for determination of
$\tilde{\rm g}$ (see also \cite{BV_CUBU_2}, page 16). One of them
is to write the equation $\Delta_{\tilde{\rm g}}\Re\hat w^f=0$ for
a rich enough set ${f}={f}_1,\dots{f}_n$ in coordinates and then
use these equations as a system for finding $\tilde{\rm g}_{ij}$
(up to a smooth functional factor). Also, it is easily seen that
one can choose the metric $\tilde{\rm g}$ to be $\tau'$--
invariant, i.e., obeying $\tau'_*\tilde{\rm g}=\tilde{\rm g}$.

Let such a metric $\tilde{\rm g}$ be chosen. Find a smooth
positive function $\rho$ on ${\mathbb M}^{\,'}$ provided
$\rho=\rho\circ\tau'$ and such that the length element $ds''$ of
the metric $g''=\rho\tilde{\rm g}$ at $\Gamma'_+=\Gamma$ coincides
with the (known) element $ds$.
\smallskip

{\it Step 6.}\,\,\,Passing to the factor-space $M':={\mathbb
M}^{\,'}\slash\tau'$, we get a homeomorphic copy of the unknown
original $M$. The projection $\pi':{\mathbb M}^{\,'}\to M'$ (a
copy of the unknown $\pi$) is a local homeomorphism by
construction. Endowing $M'$ with the metric
$g'=\pi^{-1}_*\tilde{\rm g}$, we obtain the manifold $(M',g')$,
which satisfies $\partial M'=\Gamma$ and $\Lambda_{g'}=\Lambda_g$
by construction, and thus solves the EIT problem.
\bigskip

It is worth noting the following. In the course of solving the
problem by this procedure, the observer operates not with the
attributes of $M$ themselves (which is impossible in principle!),
but creates their copies, and recovers not the original $M$, but
its relevant copy $M'$. This view of what is happening is fully
consistent with the philosophy of the BC-method in inverse
problems \cite{BCald,B UMN}: the only reasonable understanding of
`to restore unreachable object' is to construct its (preferably,
isomorphic) copy.

\end{document}